\documentstyle[12pt]{article}
\begin{document}
{}~~\hfill{UICHEP-TH/92 - 9}\\
\hspace*{\fill {PRA: ~~AT4643}}\\
\hspace*{\fill{\footnotesize{draft of \today }}}\\

{}~~\\

\centerline{\large\bf Bound States in the Continuum} ~~\\
\centerline{\large\bf from Supersymmetric Quantum Mechanics} \vspace{.4in}

\centerline{\bf J. Pappademos, U. Sukhatme, and A. Pagnamenta}
\centerline{\bf Department of Physics}
\centerline{\bf University of Illinois at Chicago}
\centerline{\bf Chicago, IL ~ 60680}
\vspace{.5in}

\underline{\bf Abstract}\\

Starting from a potential with a continuum of energy eigenstates, we show how
the methods of supersymmetric quantum mechanics can be used to generate
families of potentials with bound states in the continuum [BICs]. We also find
the corresponding wave functions. Our method preserves the spectrum of the
original potential except it adds these discrete BICs at selected energies.
Specifically, we compute and graph potentials which have bound states in the
continuum starting from a null potential representing a free particle and from
both the attractive and the repulsive Coulomb potentials.\\

{}~~\\
{}~~
\thispagestyle{empty}

\vfill\hspace{\fill}{\tiny bicus.tex}
\newpage

\centerline{\bf I. Introduction} ~~\\

In 1929, Von Neumann and Wigner \cite{vn-wig} realized that it was possible to
construct potentials which have quantum mechanical bound states embedded in the
classical energy continuum (BICs). Further developments, by many authors [2-6]
have produced more examples and a better understanding of the kind of potential
that can have such bound states, although there is not as yet a fully
systematic approach. These authors have also suggested possible applications to
atoms and molecules. Furthermore, these works have shown that such BICs appear
mainly in certain oscillatory potentials whose envelopes fall off fast enough
to lead to normalizable wave functions but sufficiently slowly, such that the
different maxima are able to conspire to create a captive state. Friedrich and
Wintgen \cite{fw} have given the example of two conspiring resonances and also
of a hydrogen atom in a uniform magnetic field. Robnik \cite{robnik} has shown
in a similar way that a simple separable Hamiltonian can develop bound states
in the continuum. In his examples too, coupled channels are responsible for the
creation of the BIC. Such a BIC is a very fragile structure -  a small
perturbation of the potential transforms it into a decaying resonance.
Nevertheless, Capasso et al. \cite{capasso} have recently reported direct
evidence for BICs by constructing suitable potentials using semiconductor
heterostructures grown by molecular beam epitaxy. Finally, it is interesting to
note that BICs  have found their way into a text \cite{ball}, illustrating for
students the surprising possibility of the existence of quantum mechanical
bound states in the classical continuum.\\

Recently, extensive work has been devoted to generating isospectral potentials
by the methods of supersymmetric quantum mechanics (SUSYQM) [9 - 14]. Starting
from the Schr\"odinger equation for a potential, whose ground state wave
function is known, this method permits one to generate families of new
potentials, which may look quite different from the original one, but have
exactly the same spectrum [12-14]. These methods are based on procedures
invented by  Darboux \cite{darboux} and generalized by Crum \cite{crum}. In
this paper we extend the usual SUSYQM formalism for obtaining isospectral
potentials and  apply it to potentials with a  continuum of scattering states
to generate new potentials with bound states in the continuum. We show that,
while the wave functions in the continuum of the original potential are
non-normalizable, the ones generated by SUSYQM are normalizable thus
representing a bound state. In particular, we construct one-parameter and
two-parameter families of supersymmetric partner potentials with one and two
bound states in the continuum.\\

In Section II, we present the development of the one-parameter family in SUSYQM
to generate  potentials with a single bound state in the continuum. The
procedure is valid for any spherically symmetric potential V(r) which vanishes
as $ r \rightarrow \infty $. We illustrate our procedure with two explicit
examples: (A) free particle $V \equiv 0$ on the half line; (B) attractive and
repulsive Coulomb potentials. In Section III, we generalize the SUSYQM method
to the two-parameter family of isospectral potentials and show that it
generates two bound states in the continuum. Section IV contains a summary of
our results. The method can readily be extended to an arbitrary number of
BICs.\\

{}~~\\ \centerline{\bf II. The One Parameter Family of BICs.}\\

The radial s-wave Schr\"odinger equation for the reduced wavefunction u(r) (in
units where $ \hbar$ =2m=1) is \\ \begin{equation} -u''+V(r)~ u(r) = E~ u(r),
\end{equation} where we have scaled the energy and radial variables such that
all quantities are dimensionless. A prime denotes differentiation with respect
to r. For any potential which vanishes at infinity, Eq. (1) has a classical
continuum of positive energy solutions which are clearly not normalizable.\\

Using the formalism of SUSYQM and the Darboux \cite{darboux} procedure for
deleting and then reinstating the ground state $u_o(r)$ of a potential V(r),
one can generate a family of potentials $\hat{V}(r;\lambda )$ which have the
same eigenvalues as V(r). These isospectral potentials are labeled by a real
parameter $\lambda$ which lies in the ranges $\lambda > 0$ or $\lambda < -1$.
The isospectral potential $\hat{V}(r;\lambda )$  is given  in terms of the
original potential V(r) and the original ground state wave function $u_0(r)$ by
[12,14] \begin{equation} \label{vhat}  \hat{V}(r;\lambda)=
V(r)-2[ln(I_0+\lambda)]''=
V(r)-\frac{4u_0u_0'}{I_0+\lambda}+\frac{2u_0^4}{(I_0+\lambda)^2},\end{equation}
where \begin{equation} \label{i0} I_0(r) \equiv \int_0^r u_0^2(r')
dr'.\end{equation}

Let us recall the main steps in obtaining Eq. (\ref{vhat}). First, one writes
the Hamiltonian $H=-\frac{d^2}{dr^2} + V(r)$ in factorized form $H=A^{\dagger}
A$, with the operators $$A=\frac{d}{dr}+W(r), ~~~ A^{\dagger} =
-\frac{d}{dr}+W(r).$$ The superpotential is given by $W(r)=-u_0'/u_0.$ The
supersymmetric partner Hamiltonian is
$$H_+=AA^{\dagger}=-\frac{d^2}{dr^2}+V_+(r), $$ where
$$V_+(r)=W^2+W'=V(r)-2\left( \frac{u_0'}{u_0}\right) '.$$ If the potential
V(r) has eigenfunctions $u_n(r)$ at energies $E_n$, then the SUSY partner
potential $V_+(r)$ has the same energy eigenvalues as V(r) with eigenfunctions
$Au_n(r),$ except that there is no ground state at E=0 since $Au_0(r)=0.$ This
is the standard procedure for deleting the ground state and obtaining
$V_+(r).$
To re-insert the ground state, one asks for the most general superpotential
$\hat{W}(r)$ such that $$V_+(r)=\hat{W}^2+\hat{W}'$$ and this can be
shown to be \cite{nieto} $$\hat{W}(r;\lambda )=
W(r)+\frac{d}{dr}ln~(I_0(r)+\lambda),$$
with $I_0$ given in Eq. (\ref{i0}). Thus the entire family of potentials
$\hat{V}(r;\lambda)=\hat{W}^2(r;\lambda)-\hat{W}'(r;\lambda) $ has
the same supersymmetric partner potential $V_+(r)$
obtained by deleting the ground state.

In all previous work, $u_0$ was taken to be the nodeless, normalizable ground
state wave function of the starting potential V(r). However, for the purposes
of this paper, we can generalize the above equations to the case where $u_0(r)$
is any solution of \\ Eq. (1) with arbitrary energy $E_0.$ If $u_0(r)$ has
nodes, this leads to singular superpotentials and to singularities in the
partner potential $V_+(r).$ However, when the original state at $E_0$ is
re-inserted, the resulting family of potentials  $\hat{V}(r;\lambda)$ is free
of singularities \cite{ps}.   Our results are best summarized in the following
statement:\\

{\bf Theorem:} Let $u_0(r)$ and $u_1(r)$ be any two nonsingular solutions of
the \\Schr\"odinger equation for the potential V(r) corresponding to
arbitrarily selected energies $E_0$ and $E_1$ respectively. Construct a new
potential $\hat{V}(r;\lambda)$ as prescribed by Eq. (\ref{vhat}). Then, the two
functions \begin{equation}  \hat{u}_0(r;\lambda)=\frac{u_0(r)}{I_0+\lambda},
\label{u0hat}\end{equation} and \begin{equation}  \hat{u}_1(r;\lambda)=
(E_1-E_0)u_1 + \hat{u}_0 ~W(u_0,u_1),\label{u1hat} \end{equation}

\noindent [where W denotes the Wronskian, $W(u_0,u_1) \equiv u_0u_1'-u_1u_0'$]
are solutions of the Schr\"odinger equation for the new potential
$\hat{V}(r;\lambda)$ corresponding to the same energies $E_0$ and $E_1$.\\

While the new potential in Eq. (\ref{vhat}) and the new wave functions in Eq.
(\ref{u0hat}) were  originally inspired by SUSYQM, the easiest proof of the
above theorem  is by direct substitution. One simply computes
$-\hat{u}_i''+\hat{V}(r;\lambda)\hat{u}_i$ (i=0,1), with the wave functions
$\hat{u}_i$ given in the theorem. After straightforward but tedious algebraic
manipulations, one gets $E_i\hat{u}_i$, thus establishing the theorem. The
algebra is considerably simplified by using the following identity for the
Wronskian of two solutions of the Schr\"odinger equation: \begin{equation}
\frac{d}{dr} W(u_0,u_1) = (E_0-E_1) u_0u_1.\label{wder} \end{equation} In the
present work, we will take $u_0$ to be a scattering solution at a positive
energy $E_0=k^2$ of a potential V(r) which vanishes at r=$\infty$. Taking
$u_0(r=0)=0$ satisfies one of the required boundary conditions, but clearly
$u_0$ oscillates as $r \rightarrow \infty$ and has an amplitude which does not
decrease. Consequently, the integral $I_0(r)$ in Eq. (\ref{i0}) grows like r at
large r and $\hat{u}_0$ is now square integrable for $\lambda >0$, while the
original wave function $u_0$ was not. Therefore, we see that all the potentials
$\hat{V}(r;\lambda)$ have a BIC with energy $E_0$. Note from Eq. (\ref{u0hat})
that $\hat{u}_0$ has the same zeros as the original $u_0$. At zeros of $u_0$,
$\hat{V}(r;\lambda)$ and $V(r)$ are equal. On the other hand, all other
oscillatory solutions to $V(r)$ get transformed into oscillatory solutions to
$\hat{V}(r;\lambda)$ with the same energy. In particular, note that
$\hat{u}_1(r;\lambda)$ remains a non-normalizable scattering solution of the
corresponding Schr\"odinger equation. \\

We note that the new potential $\hat{V}(r;\lambda)$ in Eq. (\ref{vhat}) and the
BIC at energy $E_0$ are formed using the corresponding wave function $u_0(r).$
Any other state, say $u_1(r)$, is transformed into a solution of the new
Schr\"odinger equation by the operation given in Eq. (\ref{u0hat}) which
involves both $u_0$ and $u_1$. The central column of the table  gives a
convenient overview of the relationship of the potentials V and $\hat{V}$ and
the solutions of the corresponding Schr\"odinger equations.\\

We now give two examples to explicitly illustrate how one applies the above
procedure to obtain potentials possessing one BIC.\\

\centerline{\bf A. Free Particle on the Half Line.} ~~\\ Here, we consider the
case $V\equiv 0$, the free particle on the half line $0\le r < \infty$. We
choose $u_0=sinkr$, the spherical wave solution, corresponding to energy
\\$E_0=k^2>0,$ which vanishes at $r=0.$ The integral $ I_0$ given in
Eq.(\ref{i0}) becomes \begin{equation} \label{i0sp} I_0=[2kr-sin(2kr)]/(4k).
\end{equation} We observe that $ I_0 \rightarrow r/2 $ as $ r \rightarrow
\infty $. ~~\\

The potential family $\hat{V}$, defined in Eq.(\ref{vhat}) becomes

\begin{equation} \label{v0hsp} ~~~ \hat{V}(r;\lambda)
=\frac{32~k^2~sin^4kr}{D_0^2}-\frac{8~k^2sin(2kr)}{D_0} \end{equation} with
\begin{equation} \label{d0sp}~~~ D_0(r;\lambda)=2kr-sin(2kr)+ 4k\lambda.
\end{equation} $\hat{V}$ has a BIC at energy $E_0=k^2$ with wave function
\begin{equation}  \label{u0hsp} ~~~ \hat{u}_{0}(\lambda)=4k~sinkr/D_0.
\end{equation}

For special values of the parameters k and $\lambda$, the potential $ \hat{V}$
and its BIC wave functions are shown in Figs. 1a and 1b. The original null
potential has now become an oscillatory potential which asymptotically has a
1/r envelope. The new wave function at $ E_0=k^2$ also has an additional
damping factor of 1/r which makes it square integrable. As $ u_{0}$ appears in
the numerator of $ \hat{V}$, Eq. (\ref{vhat}), every node of $\hat{u}_0$ is
associated with a node of $\hat{V}$ but not every node of $\hat{V}$ produces a
node of $\hat{u}_0$. The value of the eigenenergy $ E_0$ clearly is above the
asymptotic value, zero, of the potential. Evidently, the many oscillations of
this potential, none of them able to hold a bound state, conspire in such a way
as to keep the particle trapped.\\

The parameter $\lambda$ which appears in the denominator function
$D_0(r;\lambda)$ plays the role of a damping distance; its magnitude indicates
the value of r at which the monotonically growing integral $I_0$ becomes a
significant damping factor, both for the new potential and for the new wave
function. This is illustrated graphically in Figs. 1a and 1b which are drawn
for very different values of $\lambda$. [Note that the wave functions
shown in the figures are not normalized].
The parameter $\lambda$ must be
restricted to values greater than zero in order to avoid infinities in
$\hat{V}$ and in the wave functions. In the limit $\lambda \rightarrow \infty$,
$\hat{V}$ becomes identical to V.\\

{}~~\\

\centerline{\bf B. Coulomb Potential} ~~\\ Starting from the potential $ V =
Z/r,$ for either positive or negative values of Z, one can easily construct the
one-parameter family of isospectral potentials possessing a normalizable
positive energy wave function. Here the unbound, reduced l = 0 wave function
satisfies the Schr\"odinger equation Eq.(1), which can be written in standard
form \begin{equation} \label{couleq} ~~~u_{0}''+(1-2\frac{\tilde
{\eta}}{\rho})u_{0}=0\end{equation} with $ \rho =\sqrt{E} r$ and
$ \tilde {\eta} =Z/2 \sqrt{E}.$ \\

For both positive and negative $ \tilde {\eta}$, the solutions involve
confluent hypergeometric functions which in the asymptotic limit approach sine
waves phase-shifted by a logarithmic term. Useful expressions for these
solutions in the regions near and far from the origin are available in the
literature \cite{as,nbs}. Stillinger and Herrick \cite{st-he}, following the
method of Von Neumann and Wigner \cite{vn-wig}, have constructed BIC potentials
and wave functions for the case of the repulsive Coulomb potential. Here we use
our theorem to construct a one-parameter family of isospectral potentials
containing a BIC. The procedure is the same for both positive and negative Z;
the only difference being in the sign of $ \tilde {\eta}$. The formal
expressions for the BIC potentials and wave functions have been given above,
Eqs.(\ref{vhat}) and (\ref{u0hat}), in terms of $u_{0}$.\\

For both the attractive and the repulsive Coulomb potentials, the positive
energy solution of Eq.(\ref{couleq}) can be written in the usual form
\cite{as,nbs,st-he} as the real function \begin{equation}
u_{0}(\rho)=C_0(\tilde {\eta})~e^{-i\rho}M(1-i\tilde {\eta},2,2i\rho),
\end{equation} where \begin{equation} C_0(\tilde {\eta})=(e^{-\pi\tilde
{\eta}/2})\mid\Gamma(1+i\tilde {\eta})\mid \end{equation} and M(a,b,z) is
Kummer's function. Using tabulated expressions for the Coulomb wave functions
\cite{nbs} and doing the integral for $ I_0$ numerically, we have obtained the
BIC wave functions for representative values of $\lambda.$ The corresponding
one-parameter family of potentials obtained by the SUSY procedure is given in
Eq. (\ref{vhat}) with $ V_0~=~Z/r$.\\

The results are displayed in Figs. 2 and 3. Fig. 2a shows the BIC partner to
the attractive Coulomb potential for $\lambda$ = 1, k = 1, and Z = --2.  Fig.
2b shows the (unnormalized)
wave function of the bound state in the continuum for this
potential at $E_0$ = k$^2$. Fig. 3a shows the BIC partner potential to a
repulsive Coulomb potential for $\lambda$ = 1, k = 1, Z = 6, while Fig. 3b
shows the corresponding wave functions. For comparison the original Coulomb
potentials and wave functions are also shown dotted. It is seen that the
potential which holds a bound state of positive energy shows an oscillatory
behavior about the Coulomb potential, $ V_C$, as is also evident from the form
of Eq.(\ref{vhat}) for $\hat{V}$. Since the oscillating component vanishes
whenever $u_0$ vanishes, we have $\hat{V}~=~V$ at each node of $u_0$. Compared
to the original, unnormalizable wave function, the BIC wave function in both
cases shows a damped behavior due to the denominator function. This is also
seen in the figures. \\

A similar behavior is also expected for other radially symmetric potentials
with a continuous spectrum of positive eigenvalues. For one-dimensional
potentials, the situation is not so clear cut. Our method works for the Morse
potential which is steeply rising on the negative x-axis with correspondingly
damped wave functions. It also works for the case of a particle in a constant
electric field for similar reasons. For potentials, such as V(x)=$ - V_0
{}~sech^2(x)$, the integral I$_0$ Eq.(\ref{i0}) is not convergent if the
starting
point is chosen at $ -\infty $, and it gets negative contributions if the
starting point is selected at finite x-values. This leads to a vanishing
denominator function in the expressions for some wave functions which makes
them unacceptable.\\

\centerline{\bf III. The Two Parameter Family of Potentials} ~~\\ In Section
II, we have seen that a straightforward procedure exists using the SUSY
technique for generating a completely isospectral one-parameter family of
potentials and that these potentials have a bound state in the continuum if we
select as a starting point a positive energy solution of the Schr\"odinger
equation for any potential~$ V$(r). We now show how this procedure can be
extended to construct two-parameter families which contain two BICs.\\

In constructing the new wave functions for the one-parameter family, Eq.
(\ref{vhat}), we observe that the denominator function given in Eq.
(\ref{u0hat})  was all that was needed to create the BIC, while the operation
in Eq. (\ref{u1hat}) ensured that the wave functions for all the other states,
there represented by $\hat{u}_1$, are a solution to the new potential. Note
again, there is nothing special about the ordering of the two energy values nor
the relative magnitude of $E_0$ and $E_1$, therefore we can repeat this
procedure by applying the theorem to the wave functions and the potential of
the one-parameter family, but this time we transform the state at $E_1$ into a
BIC. The state at $E_0$, which already is a BIC, is transformed in the step of
Eq. (\ref{u1hat}), suitably modified, to become a solution to the new
potential. In this way we obtain the two parameter family of potentials

\begin{equation} \label{v2hat} \hat{\hat{V}}(r;\lambda ,\lambda_1)
=\hat{V}-2[ln(\hat{I}_1+\lambda_1)]'' =
\hat{V}-\frac{4\hat{u}_1\hat{u}_1'}{\hat{I}_1+\lambda_1}+
\frac{2\hat{u}_1^4}{(\hat{I}_1+ \lambda_1)^2}  \end{equation} with the
solutions of the corresponding Schr\"odinger equation \begin{equation}
\label{u02hat} \hat{\hat{u}}_0= (E_0-E_1)\hat{u}_0 + \hat{\hat{u}}_1
{}~W(\hat{u}_1,\hat{u}_0), \end{equation} \begin{equation} \label{u12hat}
\hat{\hat{u}}_1= \frac{1}{\hat{I}_1+\lambda_1}\hat{u}_1, \end{equation} and
\begin{equation} \hat{I}_1 \equiv \int_0^r \hat{u}_1^2(r')dr'.\end{equation}

\noindent The precise relationship of the new potential and its wave functions,
which are now both BICs, is illustrated in the last column of the table.\\

While the compact form of Eqs. (\ref{v2hat} - \ref{u12hat}) explicitly shows
the method of construction, it is useful to observe that the integral
$\hat{I}_1$ can be conveniently re-cast into a simpler form which contains
integrals of the form \begin{equation} \label{ii} I_i = \int_0^r
u_i^2(r')dr',\end{equation} involving the original wave functions only. Making
use of Eq. (\ref{u1hat}) for $\hat{u}_1$, we get \begin{equation} \hat{I}_1 =
\int_0^r \left[(E_1-E_0)^2 u_1^2 +\frac{u_0^2 W^2}{(I_0+\lambda )^2}
+2(E_1-E_0)\frac{u_0u_1}{(I_0+\lambda )}W  \right]dr'. \label{iu1hat}
\end{equation} The second term is integrated by parts as \begin{equation}
\int_0^r\frac{u_0^2}{(I_0+\lambda )^2} W^2(r') dr' =
\left. \frac{-W^2}{I_0+ \lambda }
\right|_0^r +\int_0^r \frac{2WW'}{(I_0+\lambda )}dr' . \end{equation}
We
now use Eq.~(\ref{wder}) for the derivative of a Wronskian of two solutions of
the Schr\"odinger equation to rewrite the second term and observe, that it
exactly cancels the last term in Eq. (\ref{iu1hat}). We therefore have
\begin{equation} \hat{I}_1 (r) = \frac{-W^2(r)}{I_0+\lambda } +(E_1-E_0)^2
{}~I_1(r).\end{equation} Here we have made use of the fact that our boundary
conditions imply that W(0)=0.\\

As an example, we evaluate the two-parameter potential \begin{equation}
\hat{\hat{V}}=V-2 \left[ ln \left\{ (I_0+\lambda )
{[(E_1-E_0)^2I_1-\frac{W^2(r)}{I_0+\lambda } +\lambda_1]}  \right\} \right]
''. \end{equation}

The argument of the logarithm can be rewritten as \begin{equation} (E_1-E_0)^2
I_0I_1 - W^2(r) +\lambda \lambda_1 +\lambda (E_1-E_0)^2I_1 +\lambda_1 I_0.
\label{arg} \end{equation}

We happen to have transformed first the state at energy $E_0$ into a BIC and
then, in the second step, the state at $E_1,$ which introduced the parameters
$\lambda$ and $\lambda_1 .$ Let us now consider applying our procedure in the
reverse order, that is let us first transform the state at energy $E_1$ into a
BIC and then the state at energy $E_0$, producing the parameters $\mu$ and
$\mu_1$. For this situation, the argument corresponding to Eq.~(\ref{arg}) is
\begin{equation} (E_1-E_0)^2 I_0I_1 - W^2(r) +\mu \mu_1 +\mu_1 (E_1-E_0)^2I_0
+\mu I_1. \label{arg2} \end{equation} Clearly, one expects symmetry. This is
guaranteed if the parameters are related by 	\begin{equation} \mu =\lambda
(E_1-E_0)^2 \end{equation}        \begin{equation} \mu_1 =\lambda_1 /
(E_1-E_0)^2. \end{equation} This also leads to the same two-parameter
wave functions.  We also note that transforming any state twice by Eq.
(\ref{u0hat}) does not create a second denominator or anything else new, but
simply changes the value of the parameter $\lambda$ as shown in ref.\cite{ks}.
Finally, relation (\ref{u02hat}) ensures that all other eigenstates will be
solutions to the new potentials.

Figures 4 and 5 illustrate the appearance of potentials and wave functions for
representative choices of $ \lambda$ and $ \lambda_1$. Clearly, various choices
of the parameters $ \lambda$ and $ \lambda_1$ lead to quite different looking
potentials $ \hat{\hat{V}}$. We note in Fig. 5 that, as $\lambda_1$ grows
large, the
two-parameter potential approaches the shape of the one-parameter family shown
in Fig. 1 and as discussed analytically above. For easy comparison with other
works, we have chosen $k_1=2k$ in the figures, however, the ratio of $E_1$ to
$E_0$ need not be integral.\\

\centerline{\bf IV. Summary} ~~\\ We have demonstrated how the SUSY method,
originally conceived for discrete spectra, can be generalised for the
construction of BICs. We were able to show how to generate a one-parameter
family of potentials which possess a localized positive energy state, starting
from a potential V(r) which has a continuum of positive energy states. The only
requirement  that V(r) must satisfy in order that it have such a continuum is
that it approaches a constant as r $ \rightarrow \infty$. Then the solution of
the Schr\"odinger equation with the potential V(r) is, oscillatory at large r,
which we can take to be of the form sin(kr). Therefore the integral $ I_0$
Eq.(\ref{i0}), will be of the form of a constant plus $ \int_{r_0}^r
{}~sin^2(kr')~dr'$, where $ r_0$ can always be found such that, for $ r >
{}~r_0$
the solution of the Schr\"odinger equation is approximately proportional to
sin(kr). This means that $ I_0 = c_1 +r/2 + sin(kr)/(4k)$, where c$_1$ is a
constant. Therefore, $\hat{u}_0$, Eq.(\ref{u0hat}) will vanish at large r as
1/r, making it a normalizable state. Thus our procedure for constructing a BIC
from an initial potential V(r) is actually valid for any spherically
symmetric potential which approaches a constant as r $
\rightarrow \infty$. The situation is more complex for one-dimensional
potentials as discussed in the text. The SUSY procedure has in common with the
original Von Neumann-Wigner \cite{vn-wig} method that it makes the wave
functions normalizable by generating a denominator function which grows with r
as $ r \rightarrow \infty$. In the case of V=0, our denominator function,
containing I$_0$ is a special case of the form used by Von Neumann and Wigner.
We illustrated the one-parameter method for two interesting and analytically
solvable cases: V(r)= 0, the free particle, and V(r)=Z/r, the Coulomb
potential. The procedure was readily extended to obtain two-parameter families
with two BICs at arbitrarily selected energies.\\

It is pleasure to thank Prof. W. Y. Keung for many helpful discussions and for
first bringing the existence of bound states in the continuum to our attention.
This work was supported in part by the U. S. Department of Energy under grant
DE-FG02-84ER40173.\\

\newpage {\Large \bf Figure Captions}\\

Fig. 1 shows two examples of potentials $\hat{V}(r)$ (solid) and the associated
BIC wave functions $\hat{u}_0(r)$ (dashed)  in the one-parameter family
starting from V(r)=0 for k=1.0. Fig. 1a is for small lambda ($\lambda$ = 0.5)
and Fig. 1b for large lambda ($\lambda$ = 5.0).\\

Fig. 2a shows the BIC potential (solid) derived from the attractive Coulomb
potential which is also shown for comparison (dotted). Observe how the BIC
potential oscillates around the original Coulomb potential.\\

Fig. 2b shows the corresponding BIC wave function (solid) and, for comparison,
the original Coulomb wave function (dotted). The damping of the BIC wave
function, which makes it normalizable is evident.\\

Fig. 3a shows the BIC potential (solid) derived from the repulsive Coulomb
potential which is also shown for comparison (dotted). Again, the BIC potential
oscillates around the original Coulomb potential.\\

Fig. 3b shows the corresponding BIC wave function (solid) and, for comparison,
the original Coulomb wave function (dotted). The damping of the BIC wave
function, which makes it normalizable is evident. \\

Fig. 4a shows $ \hat{\hat{V}}$, a typical member of the two-parameter family
of BIC potentials obtained from V(r)=0 for k=1.0, k$_1$=2.0 and $\lambda = 1.0,
\lambda _1 =2.0$. This
potential supports two bound states in the continuum at $ E_0 = 1 $, and at $
E_1 = 4 $. The associated wave functions are shown in Fig. 4b, the lower state
at $E_0$ dashed, the higher one at $E_1$ dotted.\\

Fig. 5a shows $ \hat{\hat{V}}$, a typical member of the two-parameter family of
BIC potentials obtained from V(r)=0 for k=1.0, k$_1$=2.0 and $\lambda = 1.0,
\lambda _1 =50$. This
potential also supports two bound states in the continuum at $ E_0 = 1 $, and
at $ E_1=4 $. The associated wave functions are shown in Fig. 5b, the lower
state at $E_0$ dashed, the higher one at $E_1$ dotted. Because in this case
$\lambda _1$ is relatively large, the potential approaches the shape of the
one-parameter potential shown in Fig. 1a. Note that both energies $E_0$ and
$E_1$ are above the maximum value of $\hat{\hat{V}}$.\\

\newpage ~~\\ {\bf Table.} \\

{}~~\\ \begin{tabular}{c c c} {\bf Potentials~~~~} &  &\\  & & \\ \fbox{{$ \bf
{}~~~V~~~$}} & \fbox{{$~\rm \bf
\hat{V}=V-2[ln(I_0+\mbox{\boldmath$\lambda$})]''$~}} & \fbox{{$~\rm \bf
\hat{\hat{V}}=\hat{V}-2[ln(\hat{I}_1+\mbox{\boldmath $\lambda_1$})]'' ~$}} \\

& & ~~\\

& & ~~\\  {\bf Wave functions} & &\\

 & & \\ $ {\bf u_1}$ & ${\bf \hat{u}_1=(E_1-E_0) u_1+\hat{u}_0 W(u_0,u_1)} $ &
$ {\bf \hat{\hat{u}}_1=\frac{1}{\hat{I}_1+\mbox{\boldmath$\lambda_1$}}
\hat{u}_1}$\\
$E_1$ \rule[.01in]{1.0in}{.02in} &     \rule[.01in]{2.1in}{.02in} &
\rule[.01in]{2.1in}{.02in}\\    &  &\\

${\bf u_{0}}$ & $ {\bf \hat{u}_0= {{1}\over{I_0+\mbox{\boldmath$\lambda$}}}
u_0}$ & $ {\bf \hat{\hat{u}}_0=(E_0-E_1)
\hat{u}_0+\hat{\hat{u}}_1W(\hat{u}_1,\hat{u}_0)}$\\  $E_0 $
\rule[.01in]{1.0in}{.02in}  & \rule[.01in]{2.1in}{.02in}  &
\rule[.01in]{2.1in}{.02in}\\

\end{tabular}

{}~~\\

{}~~\\

{\bf Table Caption.}\\

The one-parameter family of potentials $\hat{V}(r;\lambda )$
(central column) depends on the
parameter $\lambda$ and has one bound state in the continuum at energy
$E_0$ with
wave function $\hat{u}_0$. Note that all other new states represented
by $\hat{u}_1$ at $E_1$ are not normalizable. The right column shows the
two-parameter family of potentials  $\hat{\hat{V}}(r;\lambda ,\lambda_1)$,
depending on the parameters $\lambda$ and $\lambda _1$,
which now has two normalizable states $\hat{\hat{u}}_0$ and $\hat{\hat{u}}_1$
in the continuum. Both families of potentials are generated from the
non-normalizable scattering states $u_0$ and $u_1$ of the original potential
V(r) shown in the first column. Using the theorem described in the text,
in the first step one produces a BIC at energy $E_0$ and in the second
step a BIC at energy $E_1$. While it is
customary to denote the lower energy state by $E_0$, this is not necessary for
our approach; $E_0$ can also be above $E_1$.\\

\end{document}